\documentclass[12pt]{article}

\usepackage{amsmath}
\usepackage{amssymb}

\usepackage{graphicx}

\begin{document}
\begin{flushright}
\hfill{USTC-ICTS-08-23}\\
\end{flushright}
\vspace{4mm}

\begin{center}

{\Large \bf Reconstructing $f(R)$ Theory from Ricci Dark Energy}

\vspace{8mm}

{\large Chao-Jun Feng}

\vspace{5mm}

{\em
 Institute of Theoretical Physics, CAS,\\
 Beijing 100080, P.R.China\\
 Interdisciplinary Center for Theoretical Study, USTC,\\
 Hefei, Anhui 230026, P.R.China
 }\\
\bigskip
fengcj@itp.ac.cn
\end{center}

\vspace{7mm}

\noindent In this letter, we regard the $f(R)$ theory as an effective description for the acceleration of the universe
and reconstruct the function $f(R)$ from the Ricci dark energy, which respects holographic principle of quantum
gravity. By using different parameter $\alpha$ in RDE, we show the behaviors of reconstructed $f(R)$ and find they are
much different in the future.

\newpage

\section{Introduction}
The accelerating cosmic expansion first inferred from the observations of distant type Ia supernovae
\cite{Riess:1998cb} has been strongly confirmed by some other independent observations, such as the cosmic microwave
background radiation (CMBR) \cite{Spergel:2006hy} and Sloan Digital Sky Survey (SDSS) \cite{:2007wu}. An exotic form of
negative pressure matter called dark energy is used to explain this acceleration. The simplest candidate of dark energy
is the cosmological constant $\Lambda$, whose energy density remains constant with time $\rho_\Lambda = \Lambda / {8\pi
G}$ and whose equation of motion is also fixed, $w_{\Lambda} = P_\Lambda / \rho_\Lambda = -1$ ($P_\Lambda$ is the
pressure) during the evolution of the universe. The cosmological model that consists of a mixture of the cosmological
constant and cold dark matter is called LCDM model, which provides an excellent explanation for the acceleration of the
universe phenomenon and other existing observational data. However, as is well know, this model faces two difficulties,
namely, the 'fine-tuning' problem and the 'cosmic coincidence' problem. The former also states: Why the cosmological
constant observed today is so much smaller than the Plank scale, while the latter states: Since the energy densities of
dark energy and dark matter scale so differently during the expansion of the universe, why they are at the same order
today? \\

To alleviate or even solve these two problems, many dynamic dark energy models were proposed such as the quintessence
model relying on a scalar field minimally interacting with Einstein gravity. Here 'dynamic' means that the equation of
state of the dark energy is no longer a constant but slightly evolves with time. So far a wide variety of scalar-field
dark energy models has been proposed including quintessence mentioned above, k-essence, tachyons, phantoms, ghost
condensates and quintom etc..Despite considerable works on understanding the dark energy have been done, we can not
answer this question at present, because we do not entirely understand the nature of dark energy before a complete
theory of quantum gravity is established, since the dark energy problem may be in principle a problem belonging to
quantum gravity\cite{Witten:2000zk}. Actually, there is still a different way to face the problem of cosmic
acceleration. Since general relativity is only tested within solar system up to now, it is possible that the observed
acceleration is not the manifestation of another ingredient in the cosmic pie, but rather the first signal of a
breakdown of our understanding of the laws of gravitation, as stressed by Lue et al.\cite{Lue:2003ky}. From this point
of view, one may consider the modification to the Einstein-Hilbert action at larger scales with higher order curvature
invariant terms such as $R^2$, $R^{\mu\nu}R_{\mu\nu}$, $R^{\mu\nu\alpha\beta}R_{\mu\nu\alpha\beta}$, or $R\Box^kR$ as
well as nonminimally coupled terms between scalar fields and geometry (such as $\phi^2R$). These terms naturally emerge
as quantum corrections in the low energy effective action of quantum gravity or string theory \cite{Lue:2003ky,
Sandvik:2002jz}. The interesting models following this line include $f(R)$ and DGP gravity, and in this letter, we will
focus on $f(R)$ theory where the modification is a function of the Ricci scalar only.\\

Although we are lacking a quantum gravity theory today, we can still make some attempts to probe the nature of dark
energy according to some principle of quantum gravity\cite{Zhang:2006av}. It is well known that the holographic
principle is an important result of the recent researches for exploring the quantum gravity(or string
theory)\cite{Witten:2000zk}. So that the holographic dark energy model (HDE) constructed in light of the holographic
principle possesses some significant features of an underlying theory of dark energy\cite{Zhang:2006av}. Recently, Gao
et.al \cite{Gao:2007ep} proposed a holographic dark energy model in which the future event horizon is replaced by the
inverse of the Ricci scalar curvature, and they call this model the Ricci dark energy model(RDE). Of course, this model
also respect the holographic principle. \\

In this letter, we regard RDE as the underlying theory of dark energy and reconstruct the corresponding $f(R)$ theory
as an equivalent description without resorting to any additional dark energy component, namely RDE is effectively
described by $f(R)$ theory. In Section II, we will briefly review RDE  and $f(R)$ models, and reconstruct function
$f(R)$ from RDE
model in Section III. In the last section we will give some conclusions.\\

\section{Briefly Review on RDE and $f(R)$ theory}

Holographic principle \cite{Bousso:2002ju} regards black holes as the maximally entropic objects of a given region and
postulates that the maximum entropy inside this region behaves non-extensively, growing only as its surface area. Hence
the number of independent degrees of freedom is bounded by the surface area in Planck units, so an effective field
theory with UV cutoff $\Lambda$ in a box of size $L$ is not self consistent, if it does not satisfy the Bekenstein
entropy bound \cite{Bekenstein:1973ur} $ (L\Lambda)^3\leq S_{BH}=\pi L^2M_{pl}^2 $, where $M_{pl}^{-2}\equiv G $ is the
Planck mass and $S_{BH}$ is the entropy of a black hole of radius $L$ which acts as an IR cutoff. Cohen et.al.
\cite{Cohen:1998zx} suggested that the total energy in a region of size $L$ should not exceed the mass of a black hole
of the same size, namely $ L^3\Lambda^4\leq LM_p^2 $. Therefore the maximum entropy is $S^{3/4}_{BH}$. Under this
assumption, Li \cite{Li:2004rb} proposed the holographic dark energy as follows
\begin{equation}\label{li}
    \rho_\Lambda = 3c^2M_p^2 L^{-2}
\end{equation}
where $c^2$ is a dimensionless constant. Since the holographic dark energy with Hubble horizon as its IR cutoff does
not give an accelerating universe \cite{Hsu:2004ri}, Li suggested to use the future event horizon instead of Hubble
horizon and particle horizon, then this model gives an accelerating universe and is consistent with current
observation\cite{Li:2004rb, Huang:2004ai}. For the recent works on holographic dark energy, see ref.
\cite{Zhang:2007sh}. In the following, we are using units $8\pi G = c = \hbar = 1$.\\

Recently, Gao et.al \cite{Gao:2007ep} proposed a holographic dark energy model in which the future event horizon is
replaced by the inverse of the Ricci scalar curvature, and they call this model the Ricci dark energy model(RDE). This
model does not only avoid the causality problem and is phenomenologically viable, but also solve the coincidence
problem of dark energy. The Ricci curvature of FRW universe is given by
\begin{equation}\label{Ricci}
    R = -6(\dot H + 2H^2 + \frac{k}{a^2}) \, ,
\end{equation}
where dot denotes a derivative with respect to time $t$ and $k$ is the spatial curvature. They introduced a holographic
dark energy proportional to the Ricci scalar
\begin{equation}\label{Ricci DE}
    \rho_X = 3\alpha \left(\dot H + 2H^2 + \frac{k}{a^2}\right) \propto R
\end{equation}
where the dimensionless coefficient $\alpha$ will be determined by observations and they call this model the Ricci dark
energy model. Solving the Friedmann equation they find the result
\begin{equation}\label{Energy density of Ricci DE}
   \frac{\rho_X}{3H^2_0}  = \frac{\alpha}{2-\alpha}\Omega_{m0}e^{-3x} + f_0e^{-(4-\frac{2}{\alpha})x}
\end{equation}
where $\Omega_{m0} \equiv \rho_{m0}/3H^2_0$, $x = \ln{a}$ and $f_0$ is an integration constant. Substituting the
expression of $\rho_X$ into the conservation equation of energy,
\begin{equation}\label{conservation law}
    p_X = -\rho_X-\frac{1}{3}\frac{d\rho_X}{dx}
\end{equation}
we get the pressure of dark energy
\begin{equation}\label{pressure of X}
    p_X = -3H^2_0\left(\frac{2}{3\alpha}-\frac{1}{3}\right)f_0e^{-(4-\frac{2}{\alpha})x}
\end{equation}
Taking the observation values of parameters they find the $\alpha \simeq 0.46 $ and $f_0 \simeq 0.65$
\cite{Gao:2007ep}. The evolution of the equation of state $w_X \equiv p_X / \rho_X $ of dark energy is the following.
At high redshifts the value of $w_X $ is closed to zero, namely the dark energy behaves like the cold dark matter, and
nowadays $w_X $ approaches $-1$ as required and in the future the dark energy will be phantom. The energy density of
RDE during big bang nucleosynthesis(BBN) is so much smaller than that of other components of the universe ($\Omega_X
|_{1MeV}<10^{-6}\ll 0.1$ when $\alpha<1$), so it does not affect BBN procedure. Further more this model can avoid the
age problem and the causality problem.\\

Consider the modification of gravity, one can add terms like $R^2$, $R^{\mu\nu}R_{\mu\nu}$,
$R^{\mu\nu\alpha\beta}R_{\mu\nu\alpha\beta}$, $R\Box^kR$ or nonminimally coupled terms to the effective Lagrangian of
gravitational field when quantum corrections are considered. In $f(R)$ theory, the modification is adding a function of
the Ricci scalar only, and the action of it is as follows
\begin{equation}\label{action of fr}
    S = \int d^4x \sqrt{-g}\left[ f(R)+ \mathcal{L}_m \right] \, ,
\end{equation}
where $\mathcal{L}_m$ is the matter Lagrangian, and $f(R)$ is a function of $R$. Then we want to obtain the modified
Friedmann equations by varying the generalized Lagrangian. However, it is not clear how the variation has to be
performed\cite{Capozziello:2005ku}. Assuming the FRW metric, the equations governing the dynamics of the universe are
different depending on whether one varies with respect to the metric only or with respect to the metric and the
connection. These two possibilities are usually called the metric and the Palatini\cite{Palatini} approach
respectively. It is only in the case of Einstein gravity $f(R) = R$ that these two methods give the same result. The
problem of which method should be used is still a open question and a definitive answer is likely far to come.  In
ref.\cite{Capozziello:2005ku}, a method was proposed to reconstruct the form of $f(R)$ from a given Hubble parameter
$H(z)$ from observational data such as SN's Gold data\cite{Riess:1998cb} in the metric formulation. What is needed to
reconstruct $f(R)$ in their approach is an expression for $H(z)$, so we can use a $H(z)$ predicted by a given dark
energy model to determine what is the $f(R)$ theory which give rise to the same dynamics\cite{Capozziello:2005ku}. In
the following, we will follow this method to reconstruct $f(R)$
from RDE, so we use the metric formulation, although the dynamical equations are more simply in Palatini formulation.\\

Variation with respect to the metric leads to the modified Einstein equation \cite{Capozziello:2003tk}
\begin{equation}\label{modified Einstein equation}
    G_{\mu\nu} = R_{\mu\nu}-\frac{1}{2}g_{\mu\nu}R = T^{(curv)}_{\mu\nu} + T^{(m)}_{\mu\nu} \, ,
\end{equation}
where $G_{\mu\nu}$ is the Einstein tensor and
\begin{equation}\label{curvature tensor}
    T_{\mu\nu}^{(curv)} = \frac{1}{f'(R)}\left\{ \frac{1}{2}g_{\mu\nu}\left[f(R)-Rf'(R)\right]
    + f'(R)^{;\alpha\beta}\left(g_{\mu\alpha}g_{\nu\beta} - g_{\mu\nu}g_{\alpha\beta}\right) \right\}
\end{equation}
and the stress-energy tensor of matter
\begin{equation}\label{stress tensor of matter}
    T^{(m)}_{\mu\nu} = \tilde T^{(m)}_{\mu\nu} /  f'(R)
\end{equation}
with $\tilde T^{m}_{\mu\nu}$ the standard minimally coupled matter stress-energy tensor. Here and in the following, we
denote with a prime the derivative with respect to $R$. With the FRW metric, we obtain the modified Friedmann equations
\begin{eqnarray}
  H^2 + \frac{k}{a^2} &=& \frac{1}{3}\left[\rho_{curv} + \frac{\rho_m}{f'(R)}\right] \label{Fried1}\\
  2\frac{\ddot a}{a}+H^2 + \frac{k}{a^2} &=& -\left(p_{curv}+p_{m}\right) \label{Fried2} \, ,
\end{eqnarray}
where $\rho_m$ and  $p_m$ are the matter-energy density and pressure respectively, and we have defined the same
quantities for the effective curvature fluid as:
\begin{equation}\label{rho curv}
    \rho_{curv} = \frac{1}{f'(R)}\left\{ \frac{1}{2}\left[f(R)-Rf'(R)\right] - 3H\dot Rf''(R) \right\}
\end{equation}
and
\begin{equation}\label{pressure curv}
    p_{curv} = \frac{1}{f'(R)}\left\{2\frac{\dot a}{a}\dot Rf''(R) + \ddot Rf''(R)+\dot R^2f'''(R) - \frac{1}{2}\left[ f(R)-Rf'(R)\right]
    \right\} \,.
\end{equation}
Applying the Bianchi identity to eq.(\ref{modified Einstein equation}), we obtain the conservation law for the total
energy density $\rho_{tot} = \rho_{curv}+\rho_m/f'(R)$ as follows
\begin{equation}\label{conservation for tot}
    \dot\rho_{tot} + 3H(\rho_{tot} + p_{tot}) = 0 \,.
\end{equation}

In the following, we will study the case of flat universe, i.e. the spatial curvature $k=0$ with the matter as dust,
namely $p_m = 0$ and we do not consider the interaction between the matter and the curvature fluid. Thus, the matter
energy density is conserved so that $\rho_m = 3H_0^2\Omega_{m0}e^{-3x}$. \\

In fact, eq.(\ref{Fried1}),(\ref{Fried2}) and (\ref{conservation for tot}) are not independent, so we will consider
only eq.(\ref{Fried1}) and (\ref{conservation for tot}). Combine these two equations we obtain
\begin{equation}\label{combine two}
    \dot H  = -\frac{1}{2f'(R)} \biggr{\{} 3H^2_0\Omega_{m0}e^{-3x} + \ddot R f''(R)+\dot R \left[ \dot Rf'''(R) - Hf''(R)\right]
    \biggr{\}} \, .
\end{equation}
Using the relation $d/dt = H d/dx$ to replace the variable $t$ by $x$, eq.(\ref{combine two}) can be rewritten as a
third order differential equation of $f(x)$
\begin{equation}\label{equation of f}
    \mathcal{C}_3(x) \frac{d^3f}{dx^3} + \mathcal{C}_2(x) \frac{d^2f}{dx^2} + \mathcal{C}_1(x) \frac{df}{dx} =
    -3\Omega_{m0}e^{-3x} \, ,
\end{equation}
where $C_n(x)$ consists of $h(x)\equiv H(x)/H_0$ and its derivatives, where $H_0$ is the present Hubble parameter,see
Appendix A. From eq.(\ref{equation of f}), one can see that what is needed to reconstruct $f(R)$ is an expression for
$h(x)$. As a consequence, one could adopt for $h(x)$ predicted by a given dark energy model and determine what is the
$f(R)$ theory\cite{Capozziello:2005ku}. In the following, we will reconstruct $f(R)$ according to RDE.

\section{Reconstructing $f(R)$ from RDE}
From eq.(8) in \cite{Gao:2007ep}, we get
\begin{equation}\label{h for RDE}
    h^2(x) = \frac{2}{2-\alpha}\Omega_{m0}e^{-3x} + f_0e^{-\left(4-\frac{2}{\alpha}\right)x} \, ,
\end{equation}
in RDE with $h(x=0)=1$ as definition. Thus
\begin{equation}\label{boundary 4}
    \frac{2}{2-\alpha}\Omega_{m0} + f_0 = 1 \, .
\end{equation}

The differential equation (\ref{equation of f}) is so complicated that we will solve it numerically. According to
ref.\cite{Capozziello:2005ku}, the boundary conditions, i.e. the values of $f$ and its first and second derivatives
with respect to $x$ evaluated at $x=0$ are as follows
\begin{eqnarray}
  \left(\frac{df}{dx}\right)_{z=0} &=& \left(\frac{dR}{dx}\right)_{z=0} \label{boudary1}\\
  \nonumber   &&\\
  \left(\frac{d^2f}{dx^2}\right)_{z=0} &=& \left(\frac{d^2R}{dx^2}\right)_{z=0} \label{boudary2}\\
  \nonumber &&\\
  f(x=0) = f(R_0) &=& 6H^2_0\left(1-\Omega_{m0}\right)+R_0 \label{boudary3}\,.
\end{eqnarray}
These conditions are chosen on the basis of physical consideration only. Rewrite eq.(\ref{Fried1}) explicitly with
$8\pi G$ and $k=0$ as
\begin{equation}\label{Fried with G}
    H^2 = \frac{8\pi G}{3}\left[\rho_{curv} + \frac{\rho_m}{f'(R)}\right] \, .
\end{equation}
This equation shows that the function $f'(R)$ is equivalent to redefine the Newton gravitational constant $G$ as
$G/f'(R)$, that is time dependent in $f(R)$ theory. In order to be consistent with solar system experiments at $x=0$,
the effective gravitational constant $G/f'(R_0)$ must equal to $G$, thus $f'(R_0) = 1$, so we get
\begin{equation}\label{boundary 1 reason}
    f'(R_0) = 1 \rightarrow \left[\left(\frac{dR}{dx}\right)^{-1}\frac{df}{dx}\right]_{z=0} = 1
\end{equation}
which leads to eq.(\ref{boudary1}). A second condition comes from that any $f(R)$ theory must fulfill the condition
$f''(R_0) = 0$ in order to not contradict solar system tests \cite{solar test}, then it gives rise to
eq.(\ref{boudary2}). Finally, the present value of $\rho_{curv}$ in eq.(\ref{rho curv}) is
\begin{equation}\label{present rho curv}
    \rho_{curv}(x=0)=\frac{f(R_0)-R_0}{2} \,.
\end{equation}
By using the eq.(\ref{Fried1}) evaluated at present ($x=0$), and eq.(\ref{present rho curv}), we obtain the final
boundary condition eq.(\ref{boudary3}). \\

Given $\alpha=0.46$, $\Omega_{m0}=0.27$ and $h(x)$ in eq.(\ref{h for RDE}), with relation eq.(\ref{boundary 4}) and
three boundary condition eq.(\ref{boudary1})-(\ref{boudary3}), the differential equation eq.(\ref{equation of f}) can
be solved numerically. We plot the function $f(R)$ with respect to $R$ in Fig.1.  \\

\bigskip{
    \vbox{
            {
                \nobreak
                \centerline
                {
                    \includegraphics[scale=1.0]{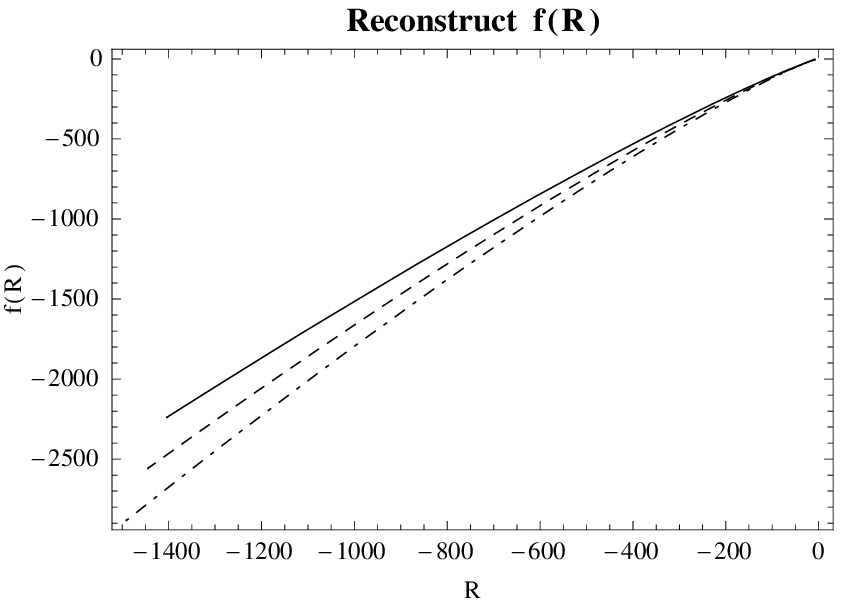}
                }
                \nobreak
                \bigskip
                {\raggedright\it \vbox
                    {
                        {\bf Figure 1.}
                        {\it Reconstructed f(R) with $0\leq z \leq 10$, where the redshift $z=e^{-x}-1$, $\Omega_{m0}=0.27$ and $\alpha = 0.46$ (solid),
                        $\alpha = 0.5$ (dashed) and $\alpha = 0.54$ (dash-dotted).
                        }
                    }
                }

            }
        }
\bigskip}

From Fig.1 one can see that for small $|R|$ (small $z$ also), the functions are distinguishable for different parameter
$\alpha$. Differences between these function $f(R)$ become significant when $|R|$ (or $z$) increases. In order to
compare with the results in \cite{Capozziello:2005ku}, we also show our results on a $lf-lR$ plane in Fig.2, where $lf
\equiv \ln(-f)$ and $lR \equiv \ln(-R)$ used in \cite{Capozziello:2005ku}, and our results are consistent with theirs.
 Moreover, Fig.1 and Fig.2 indicate the parameter $\alpha$ plays a important role
in the remote past. Actually, the value of $\alpha$ also determines the future evolution of $R$. To illustrate this, we
plot the evolution of $R$ in the future in Fig.3. As is expected, for $\alpha<0.5$, the curves indicate
$|R|\rightarrow\infty$ in the future, which is the behavior of phantom with equation of state smaller than $-1$
dominating over others. The energy density of phantom increases with time, tears apart structures and a Big Rip is
unavoidable. For $\alpha=0.5$, $|R|$ varies a little, and
the dark energy becomes more and more like a cosmological constant. For $\alpha>0.5$, $R$ will vanish in the future.\\

\bigskip{
    \vbox{
            {
                \nobreak
                \centerline
                {
                    \includegraphics[scale=0.95]{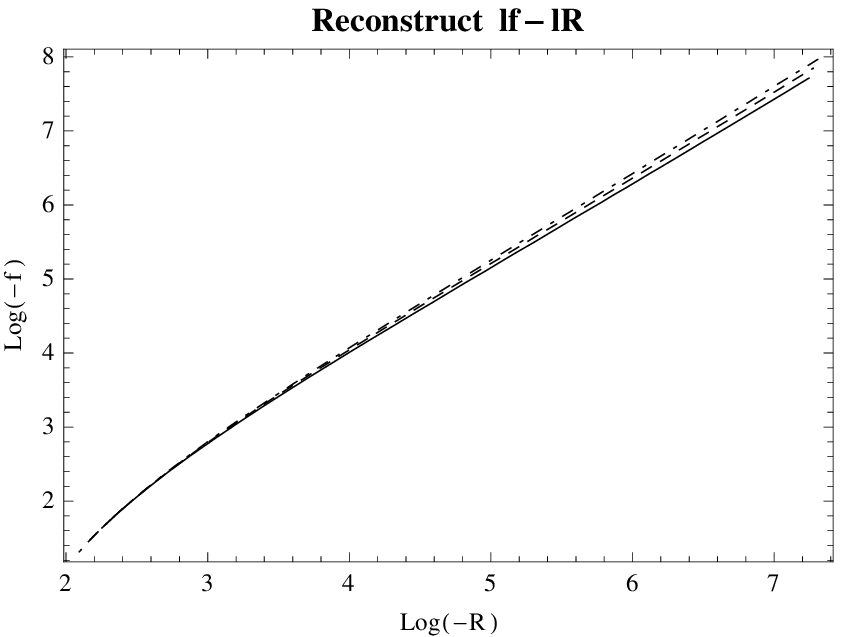}
                }
                \nobreak
                \bigskip
                {\raggedright\it \vbox
                    {
                        {\bf Figure 2.}
                        {\it Reconstructed f(R) in $lf-lR$ plane with $0\leq z \leq 10$, where the redshift $z=e^{-x}-1$, $\Omega_{m0}=0.27$ and $\alpha = 0.46$ (solid),
                        $\alpha = 0.5$ (dashed) and $\alpha = 0.54$ (dash-dotted). $lf \equiv \ln(-f)$ and $lR \equiv
                        \ln(-R)$.
                        }
                    }
                }

            }
        }
\bigskip}

\bigskip{
    \vbox{
            {
                \nobreak
                \centerline
                {
                    \includegraphics[scale=1.0]{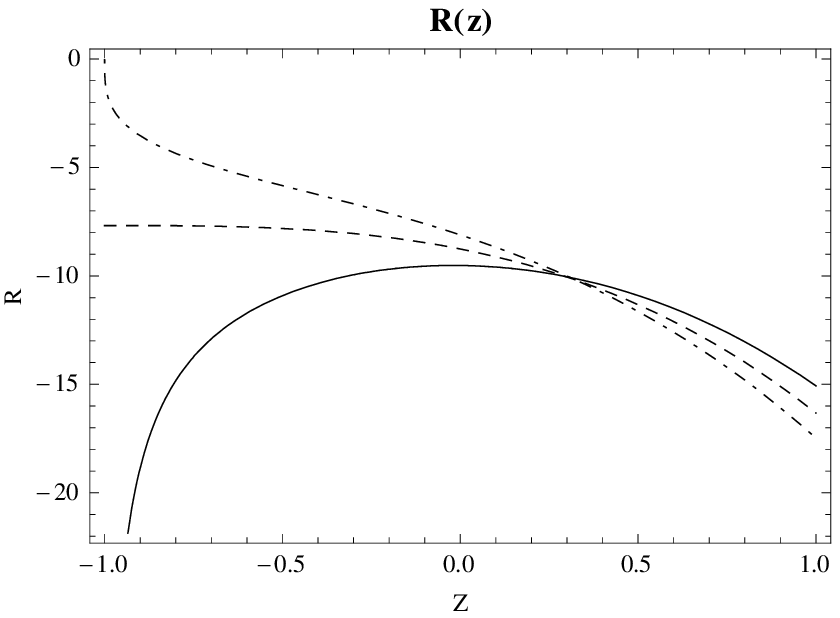}
                }
                \nobreak
                \bigskip
                {\raggedright\it \vbox
                    {
                        {\bf Figure 3.}
                        {\it The future evolution of $R$ with $0\leq z \leq 10$, where the redshift $z=e^{-x}-1$, $\Omega_{m0}=0.27$ and $\alpha = 0.46$ (solid),
                        $\alpha = 0.5$ (dashed) and $\alpha = 0.54$ (dash-dotted).
                        }
                    }
                }

            }
        }
\bigskip}

\bigskip{
    \vbox{
            {
                \nobreak
                \centerline
                {
                    \includegraphics[scale=1.0]{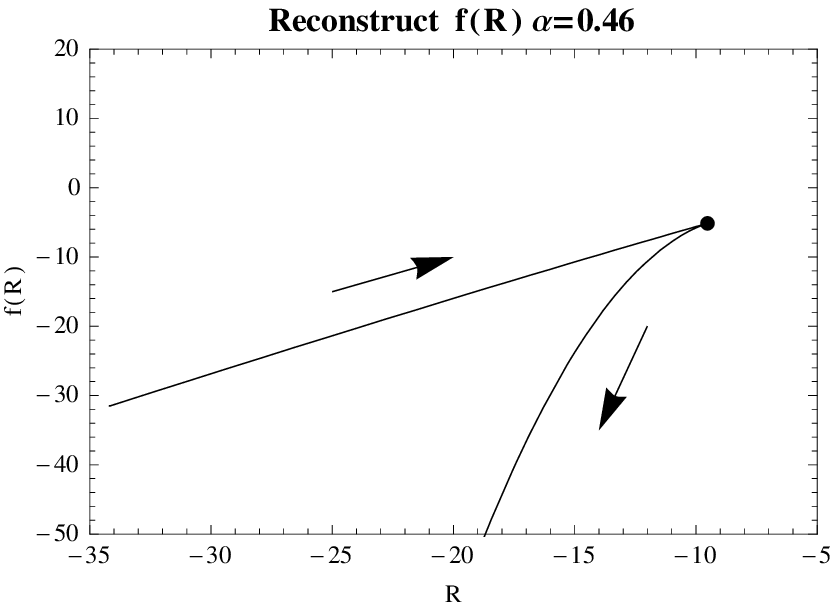}
                }
                \nobreak
                \bigskip
                {\raggedright\it \vbox
                    {
                        {\bf Figure 4.}
                        {\it Reconstructed f(R) with the redshift $z=e^{-x}-1$ from around $2$ down to $-1$, $\Omega_{m0}=0.27$ and $\alpha =
                        0.46$. The arrow denotes the decreasing direction of $z$, and the point corresponds to the
                        current value at $z=0$.
                        }
                    }
                }

            }
        }
\bigskip}

\bigskip{
    \vbox{
            {
                \nobreak
                \centerline
                {
                    \includegraphics[scale=1.0]{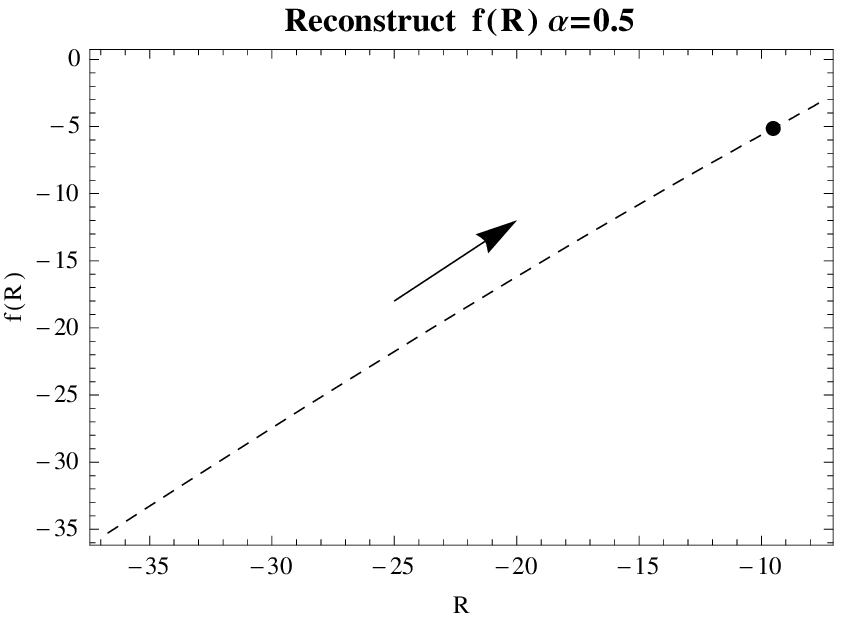}
                }
                \nobreak
                \bigskip
                {\raggedright\it \vbox
                    {
                        {\bf Figure 5.}
                        {\it Reconstructed f(R) with the redshift $z=e^{-x}-1$ from around $2$ down to $-1$, $\Omega_{m0}=0.27$ and $\alpha =
                        0.5$. The arrow denotes the decreasing direction of $z$, and the point corresponds to the
                        current value at $z=0$.
                        }
                    }
                }

            }
        }
\bigskip}

\bigskip{
    \vbox{
            {
                \nobreak
                \centerline
                {
                    \includegraphics[scale=1.0]{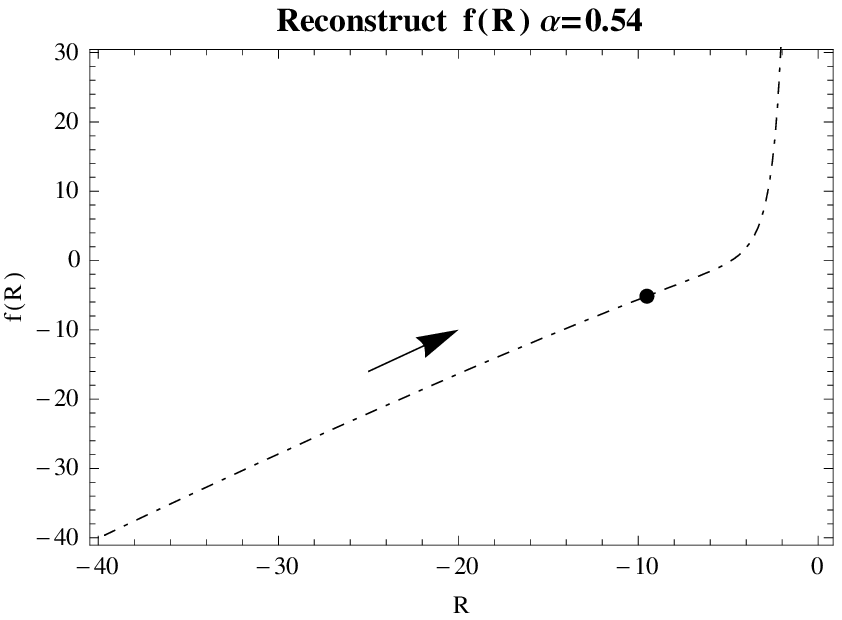}
                }
                \nobreak
                \bigskip
                {\raggedright\it \vbox
                    {
                        {\bf Figure 6.}
                        {\it Reconstructed f(R) with the redshift $z=e^{-x}-1$ from around $2$ down to $-1$, $\Omega_{m0}=0.27$ and $\alpha =
                        0.54$. The arrow denotes the decreasing direction of $z$, and the point corresponds to the
                        current value at $z=0$.
                        }
                    }
                }

            }
        }
\bigskip}

In Fig.4, Fig.5 and Fig.6, we can see that the difference is more distinctively reflected by the function $f(R)$
reconstructed according to the future evolution of RDE. For $\alpha = 0.46$, there exists a turnaround point on the
curve at present epoch ($z=0$), and the decreasing $|R|$ begin to increase at this point, which means the phantom-like
dark energy will dominate the universe. The existence of the turnaround point is a common feature for all the
phantom-dark energy models realized in $f(R)$ theory because of the competition between dark energy and matter
\cite{Wu:2007tn}. For $\alpha = 0.5$, $f(R)$ linearly depends on $R$ up to a constant, which corresponding to the de
Sitter space, where $f(R) = R + 2\Lambda$. For $\alpha = 0.54$, $f(R)$ increases from negative to positive and seems a
inverse power law dependence on $R$ in the future.

\section{Conclusions}

In conclusion, we have followed the method proposed in ref.\cite{Capozziello:2005ku} to reconstruct the function $f(R)$
in the extended theory of gravity according to the Ricci dark energy model, which respects holographic principle of
quantum gravity. We show the behaviors of $f(R)$ reconstructed with parameter $\alpha = 0.46, 0.5, 0.54$ in RDE and
find that the dependence of $f(R)$ on $R$ is different for different $\alpha$, and such a difference is much more
distinctive in the future. The basic reconstruction procedure is simply: once the function $h(x)$ given by some dark
energy model, one can solve the differential equation (\ref{equation of f}) to obtain $f$ with relation
eq.(\ref{boundary 4}) and three boundary condition eq.(\ref{boudary1})-(\ref{boudary3}). In our case, $h(x)$ is given
by RDE in (\ref{h for RDE}) and results of reconstruction are shown in Fig.1-6. The parameter $\alpha$ plays a
important role to determine the dependence of $f$ on $R$, so we hope that the future high precision observation data
may be able to
determine it and reveal some significant features of the underlying theory of dark energy.\\

It should be noted that RDE is obtained within the framework of general relativity, rather than any other extended
gravity theory such as $f(R)$ theory. What we done in this letter is to reconstruct the $f(R)$ theory to effectively
describe RDE in Einstein gravity. Whether RDE can be generalized to $f(R)$ theories is question worth further
investigation as HDE \cite{Wu:2007tn}.

\section*{ACKNOWLEDGEMENTS}
The author would like to thank Miao Li for a careful reading of the manuscript and valuable suggestions. We are
grateful to Qing-Guo Huang for useful discussions.

\section*{Appendix A}
The coefficients $\mathcal{C}_n(x)$ in eq.(\ref{equation of f}) are
\begin{equation}\label{C1}
\begin{split}
    \mathcal{C}_1 = 2h^2\left(\frac{d^2R}{dx^2}\right)^2\left(\frac{dR}{dx}\right)^{-3}
    &-\left[h^2\frac{d^3R}{dx^3}+\left(\frac{1}{2}\frac{dh^2}{dx} -
    h^2\right)\frac{d^2R}{dx^2}\right]\left(\frac{dR}{dx}\right)^{-2} \\
    &\\ &+\frac{dh^2}{dx}\left(\frac{dR}{dx}\right)^{-1}
\end{split}
\end{equation}

\begin{equation}\label{C2}
    \mathcal{C}_2 = -2h^2\left(\frac{d^2R}{dx^2}\right)\left(\frac{dR}{dx}\right)^{-2}
    +\left(\frac{1}{2}\frac{dh^2}{dx}-h^2\right)\left(\frac{dR}{dx}\right)^{-1}
\end{equation}

\begin{equation}\label{C3}
    \mathcal{C}_3 = h^2\left(\frac{dR}{dx}\right)^{-1}
\end{equation}
$R(x)$ in eq.(\ref{Ricci}) in the case of $k=0$ is
\begin{equation}\label{Ricci without r}
    R = -6\left(2h^2 + \frac{1}{2}\frac{dh^2}{dx}\right)H^2_0
\end{equation}

\end{document}